\documentclass[twocolumn,showpacs,prb,floatfix,superscriptaddress]{revtex4}

\usepackage{graphicx}
\usepackage{dcolumn}
\usepackage{amsmath}
\usepackage{amssymb}
\usepackage{amsbsy}
\usepackage[T1]{fontenc}
\usepackage{ae}
\usepackage{units}
\usepackage{color}

\usepackage{pstricks}
\usepackage{pst-node}
\usepackage{pst-blur}
\definecolor{Pink}{rgb}{1.,0.75,0.8}
\newcommand{\region}[1]{\ensuremath{\mathsf{#1}}}

\newcommand{\mmat}[1]{\ensuremath{\boldsymbol{#1}}}

\newcommand{\Tr}{\ensuremath{\mbox{Tr}}}

\begin{document}
%
%
\title{First-principles methodology for quantum transport in multiterminal junctions}

\author{Kamal K.\ Saha}
\affiliation{Computer Science and Mathematics Division, Oak Ridge National Laboratory, Oak Ridge, Tennessee 37831, USA}

\author{Wenchang Lu}
\affiliation{Center for High Performance Simulation and Department of Physics, North Carolina State University, Raleigh, North Carolina 27695-7518, USA} 

\author{J.\ Bernholc}
\affiliation{Center for High Performance Simulation and Department of Physics, North Carolina State University, Raleigh, North Carolina 27695-7518, USA} 

\author{Vincent Meunier}
\email{meunierv@ornl.gov}
\affiliation{Computer Science and Mathematics Division, Oak Ridge National Laboratory, Oak Ridge, Tennessee 37831, USA}

\date{\today}

\begin{abstract}
  We present a generalized approach for computing electron conductance
  and I-V characteristics in multiterminal junctions from first-principles. 
  Within the framework of Keldysh theory, electron transmission is 
  evaluated employing an O(N) method for electronic-structure calculations. 
  The nonequilibrium Green function for the nonequilibrium electron density 
  of the multiterminal junction is computed self-consistently by solving  
  Poisson equation after applying a realistic bias. We illustrate the 
  suitability of the method on two examples of four-terminal 
  systems, a radialene molecule connected to carbon chains and two crossed 
  carbon chains brought together closer and closer. We describe charge 
  density, potential profile, and transmission of electrons between any two
  terminals. Finally, we discuss the applicability of this technique to study 
  complex electronic devices.
\end{abstract}


\maketitle

%
%
\section{Introduction}
\label{sec:introduction}
Electron transport through molecular-scale devices has become a very exciting
research area for both experimentalists and theorists.  The main
reason for this interest originates from the possibility of extreme
miniaturization in electronic devices. In recent years, hundreds of
papers have been published to establish the connection between the
microscopic characteristics of an electronic system, such as the
atomic configuration and the electronic structure, and transport
properties such as electrical current and conductance. These results
have helped to improve the understanding of the I-V characteristics of
nanojunctions. However, the interpretation of the I-V curves in terms
of the geometry of the junction remains largely a fundamental
challenge for molecular electronics. In fact, it has not yet been
possible to establish a general theoretical model that can reliably
deal with any molecular junction of arbitrary geometry. Owing to the
complexity of the system, these studies are strongly dependent on the
existence of reliable theoretical treatments based on
first-principles approaches. In some instances, even conventional
approaches based on density-functional theory (DFT) are expected to break
down, especially in the case of weak coupling \cite{Koentopp08}.
Nevertheless, DFT is expected to provide reliable results in a 
large number of cases. 
To the best of our knowledge, all the existing approaches to date,
based on \textit{ab initio} calculations, can deal with systems limited
to two terminals only.  It is therefore of widespread interest to
develop robust computational schemes that can routinely and reliably
account for the transport mechanism in multiterminal molecular
devices. 

In his seminal work, B\"uttiker \cite{Buttiker86} developed a
conductance formula for a four-terminal system. However, that formula
was not explicitly implemented in the framework of first-principles
based calculations. Moreover, since the idea of that paper was
to propose a reliable method for voltage difference measurements,
B\"uttiker assumed that only two of the leads can carry current to and
from the sample and the two others only measure the voltage. A
  similar non-atomistic approach based on tight-binding approximation
  has been presented by Baranger {\it et al.} \cite{Baranger91}.
  Within the framework of a Luttinger liquid theory the four-terminal 
  resistance of an interacting quantum wire was studied by Arrachea 
  {\it et al.} \cite{Arrachea08}. 
  Recently, Jayasekera {\it et al.} proposed a
  four-terminal approach for magneto-transport properties based on
  R-matrix theory \cite{Jayasekera06}. However the approach is only applicable to
  two-dimensional devices, it is formulated in the framework of
  semi-empirical tight-binding, and does not include a self-consistent
  treatment of finite applied potential. Finally, a mesoscopic treatment for
  phonon-assisted current through multiterminal conductors was
  formulated by Rychkov {\it et al.}  \cite{Rychkov05}.  While these
multiterminal approaches address
  important issues, they neither treat the system in an \textit{ab
    initio} fashion, including atomistic details, nor do they account for the
  self-consistent (SC) rearrangement of electrons as the bias and the
  current increase. The importance of the self-consistency had been 
demonstrated in our previous paper \cite{Lu05}, showing that negative 
differential resistance in the I-V characteristic can only be quantitatively
studied when self-consistency is included. 

In this paper, we present a generalized approach for computing
conductance and I-V characteristics in multiterminal junctions, based
on density-functional theory. In order to take into account the
difference in electro-chemical potentials in different leads, we use
non-equilibrium Keldysh formalism. The electronic transport is
formulated in the basis of an O(N) method for electronic-structure
calculations, which is an \textit{ab initio} pseudopotential density
functional approach using a linear combination of numerical atomic
orbitals (LCAO) basis that are optimized for the problem in 
hand \cite{Fattebert00}. We apply external bias voltage through any 
lead in a realistic way and self-consistently compute the nonequilibrium
Green function for the nonequilibrium electron density of the
multiterminal junction by solving Poisson equation. One of the main
advantages of our scheme is that we can apply bias through any number
of leads and at the same time compute current between any two leads.
The method is illustrated on two prototypical four-terminal systems
(i) a radialene molecule connected to carbon chains, and (ii) two
crossed carbon chains brought together closer and closer.  We discuss
the charge density, potential profile, and transmission of electrons
between any two terminals. We also evaluate the current flowing
between them.  The algorithmic approach has been implemented on
massively parallel computer architectures, and can therefore be
applied to systems of realistic sizes.

The paper is organized as follows. The basic theory is outlined in
section~\ref{sec:theory}, sketching a nonequilibrium Green function 
formulation of the conductance calculations. The applications are discussed
in Section~\ref{sec:applications}. The scheme is applied to 
two simple systems, addressing in particular the symmetry in transmission
and effect of the different bias voltages on the I-V curves. 
The paper concludes with a summary in Section~\ref{sec:concluding-remarks}.

\section{Theory}
\label{sec:theory}
In this section, we describe the theoretical aspects of 
conductance calculations. The general system setup, the coupling to
the leads, the equilibrium and nonequilibrium density matrices, the 
implementation of the bias voltage through any number of leads, 
and the conductance formula are presented. 

\begin{figure}
  \centering
  \includegraphics[width = 0.8\columnwidth]{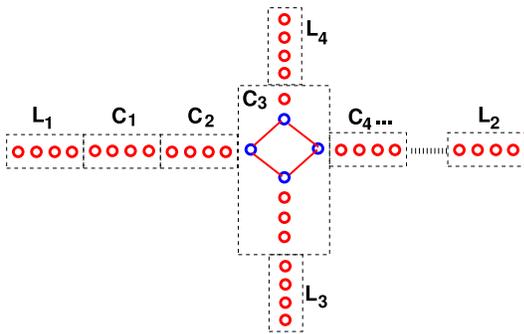}
  \caption{(Color online) Schematic diagram of a multiterminal junction. The barrier
   region is divided into $n$-number of blocks $\region{C_1, C_2, C_3,\cdots, C_n}$~.
   The leads $\region{L_1, L_2}$ are connected to blocks $\region{C_1, C_n}$,
   respectively, whereas the remaining leads $\region{L_3, L_4}$ are
   connected to block $\region{C_3}$.}
  \label{fig:sahafig1}
\end{figure}

\subsection{System setup}
We consider the prototypical multiterminal system sketched in
Fig.~\ref{fig:sahafig1}.  Two or more semi-infinite leads
$\region{L_1, L_2, L_3, \ldots}$ are coupled to a central barrier
region {\region C} with thermal reservoirs that are maintained at the
electro-chemical potentials $\region{\mu_1, \mu_2, \mu_3, \cdots}$.  Within
our approach, region {\region C} can be treated as $n$ subregions as
$\region{C_1, C_2, C_3, \ldots, C_n}$ and, in principle, we may connect
$m$-number of leads to them.  Note that an important hypothesis of the
approach (also implicit in all two-terminal approaches based on Green
function) is that there are no direct interactions between the leads and
they only interact via the barrier region. Consequently the overlap integrals
between orbitals on atoms situated in different leads take place via
the barrier region only.

\begin{figure}
  \centering
  \includegraphics[width = 0.9\columnwidth]{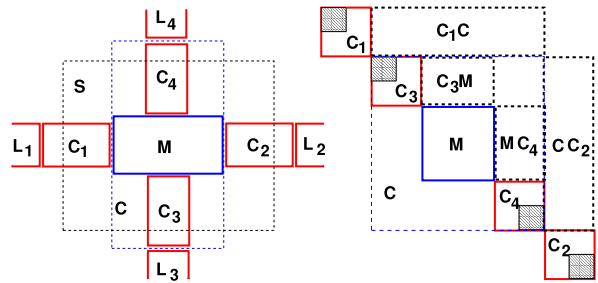}
  \caption{(Color online) (Left) Schematic of a four-terminal
  system. The leads $\region{L_1, L_2, L_3}$ and $\region{L_4}$ are 
  connected to the molecular barrier $\region{M}$ via the subregions 
  $\region{C_1, C_2, C_3}$, and $\region{C_4}$ respectively. The subregions
  $\region{C_3, C_4}$ and the molecular region $\region{M}$ are  
  considered together as a single region $\region{C}$ shown by the blue-dotted box.
  The black-dotted box shows the extended-scattering region $\region{S}$. 
  (Right) Tri-diagonal matrix representation of this system. Any matrix-element 
  in the lower off-diagonal-blocks is the complex conjugate of the 
  corresponding element in the upper off-diagonal-blocks and therefore 
  one can avoid storing the lower off-diagonal-blocks. The shaded
  blocks of the matrix are directly connected to the leads and hence
  these contain the same matrix-elements as in the respective leads.}
  \label{fig:sahafig2}
\end{figure}

For reasons of simplicity, we will be discussing a four-terminal
system as an example. However, the method can be readily generalized to any
number of electrodes. Suppose the leads $\region{L_1, L_2, L_3}$ and
$\region{L_4}$ are connected to the molecular barrier $\region{M}$ via
the subregions $\region{C_1, C_2, C_3}$ and $\region{C_4}$
respectively. At the beginning of the calculation, each subregion has
the same potential and 
charge distribution as its respective connected lead.

In order to study the transport properties, in principle, we need to invert 
an infinite Hamiltonian of the infinite system which includes all parts
of the semi-infinite leads. However, the electrons injected from the 
reservoirs move ballistically through the leads and all scattering events
only occur around the barrier region
(molecular region), called extended-scattering region $\region{S}$ 
[see Fig.~\ref{fig:sahafig2} (left)]. The potential is modified only within 
a finite region of the leads, the one being connected to
the barrier region. It is therefore sufficient to consider a finite Hamiltonian 
containing all subregions $\region{C_1, C_2, C_3, C_4}$ and the barrier 
region $\region{M}$. The Hamiltonian matrix of this system is of finite rank and 
takes the form  
\begin{equation}
  \label{eq:matrix1}
  \begin{pmatrix}
    H_\region{C_1} + \Sigma_\region{\mu_1} & 0   & V_\region{C_1 M}     & 0      & 0   \\
    0  & H_\region{C_3} + \Sigma_\region{\mu_3}  & V_\region{C_3 M}     & 0          & 0   \\ 
    V_\region{M C_1}  & V_\region{M C_3}  & H_\region{M} & V_\region{M C_4}   & V_\region{M C_2} \\ 
    0   & 0   & V_\region{C_4 M}  & H_\region{C_4} + \Sigma_\region{\mu_4}  & 0  \\ 
    0   & 0   & V_\region{C_2 M}  & 0 & H_\region{C_2}+\Sigma_\region{\mu_2}  
  \end{pmatrix},
\end{equation} 
where $H_{\region{C}_i}, H_\region{M}$ are the Hamiltonian matrices in
the $i$th-lead and the molecular barrier respectively, and
$V_{\region{C}_i\region{M}}$ is the interaction between the $i$th-lead and the
barrier $\region{M}$. $\Sigma_\region{\mu}$'s are the self-energies
that couple the scattering region to the remaining parts of the
semi-infinite leads. The Hamiltonian and the charge density matrix are assumed 
to be converged to the bulk values in the leads outside the
scattering region. For practical
calculations, this assumption is tested by including larger fractions
of the leads in the scattering region and by examining the charge
convergence during the SC iterations. Here the term
``charge convergence'' refers to the conservation of total charge in
the scattering region. The charge convergence criterion is of crucial
importance, since a failure to fulfill it would indicate bad numerical
convergence or issues with the setup of the size of the extended region.
By achieving the charge convergence, we make sure that 
all the screening takes place within the scattering region. 

The Hamiltonian matrix in Eq.~\ref{eq:matrix1} can be written explicitly
in a tri-diagonal form, as explained below. Consider the subregions $\region{C_3, C_4}$ 
and the region $\region{M}$ as a single region $\region{C}$, schematically
shown in Fig.~\ref{fig:sahafig2}. One can rewrite the above Hamiltonian
matrix as:

\begin{equation}
  \label{eq:matrix2}
  \begin{pmatrix}
    H_\region{C_1} + \Sigma_\region{\mu_1} & V_\region{C_1 C}   & 0       \\
    V_\region{C C_1}                       & H_\region{C}       & V_\region{C C_2}   \\ 
    0    & V_\region{C_2 C} & H_\region{C_2}+\Sigma_\region{\mu_2}  
  \end{pmatrix},
\end{equation} 
where
\begin{equation}
  \label{eq:matrix3}
  H_\region{C} =
  \begin{pmatrix}
    H_\region{C_3} + \Sigma_\region{\mu_3} & V_\region{C_3 M}   & 0       \\
    V_\region{M C_3}                       & H_\region{M}       & V_\region{M C_4}   \\ 
    0    & V_\region{C_4 M} & H_\region{C_4}+\Sigma_\region{\mu_4}  
  \end{pmatrix},\nonumber
\end{equation} 

We proceed with the above tri-diagonal Hamiltonian for
multiterminal calculations in the same way as in a two-terminal
case.\cite{Brandbyge02} 
The most time-consuming part is the calculation
of the Green functions, i.e., the inversion of a matrix $(\epsilon S -
H)$, where H is the Hamiltonian matrix in Eq.~\ref{eq:matrix1} and S is the
overlap matrix. For a large system, the matrix can be further reduced
to a tri-diagonal matrix with smaller blocks. Its inversion can be done
by an iterative method.
It can also be efficiently carried out using
sparse algebra.  In fact, the numerical effort to invert the matrix is
independent of the number of terminals. The matrix size (or the system
size) and its sparsity play an important role in determining the
computational cost. 

\subsection{Density Matrix}
In this subsection, we first outline the procedure adopted for
computing the density matrix for a two-terminal system \cite{Brandbyge02, Thygesen03, Nardelli99}
and then generalize it for a multiterminal system.

As explained in detail in Ref.~\onlinecite{Brandbyge02} for a
two-terminal system, one may write the density matrix as
\begin{equation}
  \label{eq:densitymatrix}
  \mmat{D}_{\nu\nu'} = \int_{-\infty}^{\infty} d\epsilon 
  \left[\rho_{\nu\nu'}^\region{L_1}(\epsilon) \, 
  n_F(\epsilon - \region{\mu_1}) + \rho_{\nu\nu'}^\region{L_2}(\epsilon) \, n_F(\epsilon
  - \region{\mu_2}) \right], 
\end{equation}
\begin{equation}
\label{eq:chargedensity}
\rho_{\nu\nu'}^{\region{L}_i}(\epsilon) = \frac{1}{\pi} \left[\mathrm{G}(\epsilon) \, \Gamma_{\region{L}_i}(\epsilon)\, 
\mathrm{G}^\dagger(\epsilon)\right]_{\nu\nu'}, i=1, 2 
\end{equation}
where $\nu$ and $\nu'$ are the indexes of localized orbitals in the extended
scattering region, ${\mathrm G}$ the Green function, and
$\Gamma_{\region{L}_i}(\epsilon) = i\left[ \Sigma_{\region{L}_i}(\epsilon) 
-\Sigma_{\region{L}_i}(\epsilon)^\dagger \right]/2$
is the coupling function for the $i$th-lead.
$\Sigma_{\region{L}_i}(\epsilon) = \left[ V\, g^i(\epsilon)\, V^\dagger \right]$
is the self-energy of the $i$th-lead that couples it to the extended-scattering region.

The density matrix given in Eq.~\ref{eq:densitymatrix} is general, i.\,e.,\ it is valid
for both equilibrium or nonequilibrium electron transport. It can be
separated into two parts  
\begin{eqnarray}
  \label{eq:densitymatrix3}
  \mmat{D}_{\nu\nu'} &=& -\frac{1}{\pi}\, \mbox{Im}\left[\int^\infty_{\mbox{EB}} d\epsilon \, 
   \mathrm{G}(\epsilon + i\delta) \, n_F(\epsilon -\region{\mu_1})\right] \nonumber \\
   &+&\int_{-\infty}^{\infty} d\epsilon \,
 \rho_{\nu\nu'}^\region{L_2}(\epsilon)\,\left(n_F(\epsilon - \region{\mu_2}) - n_F(\epsilon - \region{\mu_1})\right). 
\end{eqnarray}
where {\mbox{EB}} is the low energy bound for the valence band. The first and second parts contain the equilibrium and nonequilibrium density matrices, respectively. 

\subsection{Generalized density matrix for multiterminal junction}
Equation~\ref{eq:densitymatrix3} is now generalized to the general
multiterminal case. The generalized density matrix is now
\begin{widetext}
\begin{eqnarray*}
\mmat{D}_{\nu\nu'} &=& \sum_i \int_{-\infty}^{\infty} d\epsilon \,
\rho_{\nu\nu'}^{\region{L}_i}(\epsilon) \, n_F(\epsilon - {\region{\mu}_i}) \nonumber \\
&=& \int_{-\infty}^{\infty} d\epsilon \, \rho_{\nu\nu'}^{\region{L}_m}(\epsilon) \, n_F(\epsilon - {\region{\mu}_m}) 
 + \sum_{j\neq m} \int_{-\infty}^{\infty} d\epsilon \,
\rho_{\nu\nu'}^{\region{L}_j}(\epsilon) \, n_F(\epsilon - \region{\mu}_j) \nonumber \\
&=& \int_{-\infty}^{\infty} d\epsilon \, \left(\sum_i
\rho_{\nu\nu'}^{\region{L}_i}(\epsilon)\right) \, n_F(\epsilon - {\region{\mu}_m}) 
 + \sum_{j\neq m} \int_{-\infty}^{\infty} d\epsilon \,
\rho_{\nu\nu'}^{\region{L}_j}(\epsilon) \, n_F(\epsilon - \region{\mu}_j) 
 - \sum_{j\neq m} \int_{-\infty}^{\infty} d\epsilon \,
   \rho_{\nu\nu'}^{\region{L}_j}(\epsilon) \, n_F(\epsilon - {\region{\mu}_m}) \nonumber \\
&=& -\frac{1}{\pi}\, \mbox{Im}\left[\int^\infty_{\mbox{EB}} d\epsilon \, \mathrm{G}(\epsilon + i\delta) \, n_F(\epsilon-{\region{\mu}_m})\right] 
 + \sum_{j\neq m} \int_{-\infty}^{\infty} d\epsilon \, \rho_{\nu\nu'}^{\region{L}_j}(\epsilon) 
   \left[ n_F(\epsilon - \region{\mu}_j) - n_F(\epsilon - {\region{\mu}_m}) \right], 
\end{eqnarray*}
\end{widetext}
where energy EB is chosen to be low enough to include all of the valence bands
and $\mu_m$ is the electro-chemical potential of the $m$th-lead. In practice, each
$D_{\nu\nu'}$ is calculated separately for each $\mu_m$ being equal to the chemical 
potential of a given lead $m$. All $D_{\nu\nu'}$ to reduce the numerical 
error related to the integration.
                                                                                                                               
For $\mu_m = \mu_i$, the electro-chemical potential of the lead $i$, the density
matrix is
\begin{equation}
\label{eq:10}
\mmat{\tilde{D}}_{\nu\nu'}^i = \mmat{D}_{\nu\nu'}^i
 + \sum_{j\neq i}\Delta_{\nu\nu'}^{ij}
\end{equation}
where
\begin{eqnarray*}
\label{eq:10a}
&& \mmat{D}_{\nu\nu'}^i = -\frac{1}{\pi}\, \mbox{Im}\left[\int^\infty_{\mbox{EB}} d\epsilon \, 
\mathrm{G}(\epsilon + i\delta) \, n_F(\epsilon-\region{\mu}_{i})\right] \nonumber \\
&&\mmat{\Delta}_{\nu\nu'}^{ij} = \int_{-\infty}^{\infty} d\epsilon \, 
\rho_{\nu\nu'}^{\region{L}_j}(\epsilon)\left[ n_F(\epsilon -
\region{\mu}_j) - n_F(\epsilon - \region{\mu}_i) \right].
\end{eqnarray*}

$\mmat{D}_{\nu\nu'}^i$ and $\mmat{\Delta}_{\nu\nu'}^{ij}$ are the
equilibrium and nonequilibrium parts of the density matrix,
respectively. The integral in the first part of Eq.~(\ref{eq:10}) can
be carried out with complex contour integral technique as in
Ref.~\onlinecite{Brandbyge02}.  However, the integral in the
second part, $\mmat{\Delta}_{\nu\nu'}^{ij}$, must be calculated on
the real energy axis with a very dense mesh.  

Because of errors related to numerical integration, the computed solutions 
of Eq.~(\ref{eq:10}) will not produce exactly the same results for all $i$'s.
So, in order to minimize the error in the solutions, we compute the density matrix 
as a weighted sum of $\mmat{\tilde{D}}_{\nu\nu'}^i$ in the following way:

\begin{equation}
\label{eq:10b}
\mmat{D}_{\nu\nu'} = \sum_i w_{\nu\nu'}^i \mmat{\tilde{D}}_{\nu\nu'}^i,
\end{equation}
where
\begin{eqnarray*}
&& w_{\nu\nu'}^i = \sum_{j \ne i} \sum_{k\ne j} {\left(\Delta_{\nu\nu'}^{jk}\right)^2}/{\Delta} \nonumber \\ 
&& \mbox{and} \ \ \Delta = ({\mbox{N}-1}) \sum_{i}\sum_{j\ne i} \left(\Delta_{\nu\nu'}^{ij}\right)^2,    
\end{eqnarray*}
which satisfies $\sum_i w_{\nu\nu'}^i = 1$, with N being the number of leads. 
The weight $w_{\nu\nu'}^i$ is chosen to minimize the numerical error in the 
solution.\cite{Brandbyge02} We test the convergence by increasing the density 
of the energy mesh, thereby making sure that the integration 
yields accurate final results.

\subsection{Computation of the conductance}
\label{sec:transmission}
We apply Keldysh theory for the computation of the conductance of the
multiterminal junction. Within the `electron counting' picture of
transport, the conductance $G$ of the junction is obtained from the
transmission probabilities of all scattering channels entering from 
one lead and leaving through the other,\cite{Imry99}
\begin{equation*}
  \label{eq:9}
  G(V) = G_0 \, T(V),
\end{equation*}
where $V$ is the applied bias voltage. The conductance quantum $G_0 = e^{2} / h$ 
is the inverse von-Klitzing constant (i.\,e.  the quantum
of resistance, $R_{\mathrm{K}} \approx \unit[25.8]{k\Omega}$).  
The total transmittance $T(V)$ comprises the transmission
probabilities in the `energy window of tunneling' opened by $V$.\cite{Henk03d} 

Once the potential profile is self-consistently determined,
the transmission spectrum from leads ${\region{L}_i}$ to ${\region{L}_j}$ 
under the external applied bias, $V = \region{\mu}_i - \region{\mu}_j$, can be calculated as 
\begin{equation}
\label{eq:transmission}
T_{\region{L}_{ij}}(\epsilon,V)=\frac{2e^2}{h}\Tr\left[\Gamma_{\region{L}_i}(\epsilon)\, 
\mathrm{G}^+(\epsilon)\, \Gamma_{\region{L}_j}\, \mathrm{G}^-(\epsilon)\right],  
\end{equation}
with
\begin{equation}
\Gamma_{\region{L}_i} = i\left[\Sigma_{\region{L}_i}^{\region{C}_i\region{C}_i} 
- {\Sigma_{\region{L}_i}^{\region{C}_i\region{C}_i}}^\dagger\right]\! /2, 
\ \Sigma_{\region{L}_i}^{\region{C}_i\region{C}_i} 
= V_{\region{C}_i\region{L}_i} \; g_{\region{L}_i} \; V^\dagger_{\region{C}_i\region{L}_i}, \nonumber 
\end{equation}
where $\mathrm{G}^\pm$ are the advanced and retarded Green functions for the extended-scattering
region $\region{S}$, and $g_{\region{L}_i}$ is the surface Green function of the $i$th-lead. 

The current from the lead $\region{L}_i$ to $\region{L}_j$ through the molecular barrier is given by 
\begin{equation}
I_{\region{L}_{ij}}(V)= \int_{-\infty}^{\infty} T_{\region{L}_{ij}}(\epsilon,V)
\left[f(\epsilon-{\region{\mu}_i}) - f(\epsilon - \region{\mu}_j)\right] d\epsilon, \nonumber 
\end{equation}
where $f$ is the Fermi-Dirac distribution.

  Although the present multiterminal NEGF approach was derived and carried out 
  only within DFT, it can serve as a starting point for implementation of 
  many-body corrections at the quasi-particle \cite{Hedin65, Hybertsen88} or 
  self-interaction correction \cite{Perdew81, Toher07} levels. A time-dependent 
  formulation \cite{Kurth05} is also possible.

\begin{figure}
  \centering
  \includegraphics[width = 0.8\columnwidth]{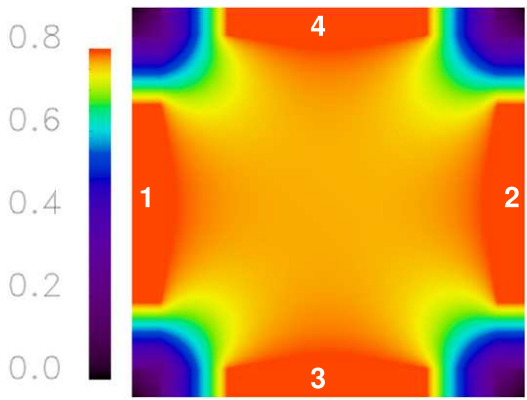}
  \caption{An initial bias profile (size 60 Bohr $\times$ 60 Bohr) for a 
  four-terminal junction to be applied 
  to the system at the beginning of a nonequilibrium calculation. For the planar
  molecules considered here, it has been generated by solving a 2D Laplace's 
  equation with appropriate boundary conditions (see text for details). An identical
  bias voltage $0.8 \, V$ is applied through all four leads $\region{L_1, L_2, L_3}$
  and $\region{L_4}$. The leads are denoted by the numbers 1, 2, 3, 4, respectively,
  and the scale bar of the potential is in eV.}
  \label{fig:sahafig3}
\end{figure}

\subsection{Finite bias}
\label{sec:bias}
In a two-terminal system, 
the initial potential for an applied bias can be simply a linear interpolation
independent of the bias between the electrodes. The situation is not as simple in 
three or four terminal system. First, one needs to make sure that the
potentials of all electrodes (outside the extended-scattering region
$\region{S}$) will be unaffected by the applied bias voltage, in other
words, the modified potential has to match at the boundary of each
electrode and the region $\region{S}$. Second, the variation of the
potential between any two electrodes through the molecular barrier has
to be continuous and uniform. Third, the electrostatic potential in 
the vacuum region between two arbitrary electrodes has to be
realistic. In order to create such an initial profile for the planar
molecules considered here,
we iteratively solve the 2D Laplace's equation (assuming the system 
is in the xy-plane) in a hypothetical system where the extended 
scattering region is empty:
\begin{equation*} 
\frac{\partial^2 V(x,y,z)}{\partial x^2} 
+ \frac{\partial^2 V(x,y,z)}{\partial y^2} = 0,
\end{equation*}

\begin{figure}
  \centering
  \includegraphics[width = 0.75\columnwidth]{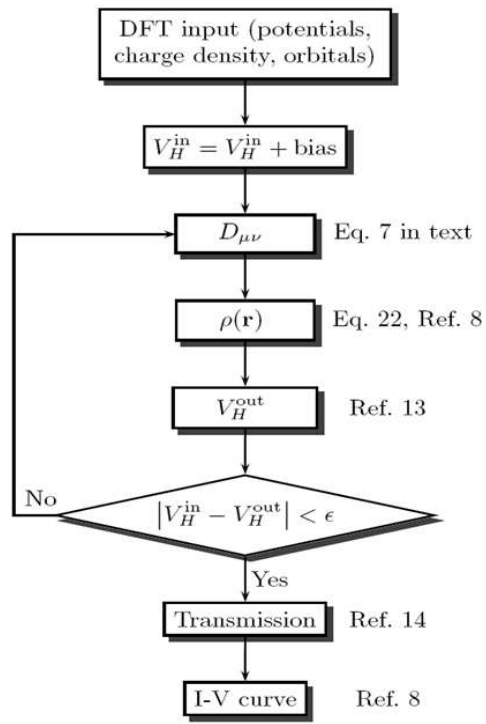}
\caption{Flowchart of the self-consistent loop used to calculate
I-V characteristics. The terminology is explained in the text.}
\label{flowchart}
\end{figure}

with the following boundary conditions: (a) the initial potential in
the scattering region is zero (or may be the same as the potentials of 
the four leads), (b) the potential in every lead is 
unchanged, (c) the potential towards the vacuum region, that
is, at the corners of the box, decays. 

Solving Laplace equation yields an initial bias-potential
profile, as shown in Fig.~(\ref{fig:sahafig3}), for a given xy-plane
of the system. In the simple of planar molecules, we repeat this image 
for the other planes of the 3D system. In more complex cases, a 3D 
solution of an approximate initial value problem would be used. 
We stress that the solution of Laplace equation is merely a
starting guess of the bias potential that is being updated via the
SC calculation. During the course of our implementation and
testing, we found that using this solution in the first iteration
significantly accelerates the convergence. It is a very
effective guess to initiating the SC iterations, as the solution to
Laplace equation is the correct one for the given setup
in absence of the central
part. Once the potential, charge density, etc. are converged, the
final result is independent of the initial guess.

\subsection{Computational Details}
\label{sec:computational}
The electronic properties of the tunnel junctions discussed in
Section~\ref{sec:applications} are obtained within the 
nonequilibrium Green function (NEGF) approach \cite{Larade01, Brandbyge02}
using a basis of optimally localized orbitals, \cite{Fattebert00, Nardelli01}
and a multi-grid approach. The \textit{ab initio} calculations for the 
leads and the molecule are performed with the O(N) method, details of 
which can be found in Ref.~\onlinecite{Fattebert00}. The exchange and
correlation terms are represented in the generalized gradient
approximation (GGA). \cite{Perdew96} 
The electron-ion interactions are described by nonlocal, ultrasoft 
pseudopotentials. \cite{Vanderbilt90} The surface Green functions are 
calculated with a transfer-matrix technique in an iterative scheme. \cite{Sancho85}
The potential and charge density in the leads are fixed to those corresponding
in the bulk material. The central conductor part includes enough ``buffer layers'' 
of the lead so that the potential and the charge density match at the 
interfaces between the conductor and leads after the SC calculations.
The Hartree potential is obtained by solving Poisson equation with
boundary conditions matching the electrostatic potentials of all the leads.
The generated SC potentials and charge density serve as inputs 
for the conductance calculations. The flowchart in Fig.~(\ref{flowchart})
explains the relations between the various steps in our algorithm.
The computations use a massively parallel real-space multigrid implementation 
\cite{Briggs96} of density-functional 
theory DFT. \cite{Kohn65} The wave functions and localized orbitals are 
represented on a grid with spacing of 0.335 Bohr. A double grid technique 
\cite{Ono99} is employed to evaluate the inner products between the nonlocal 
potentials and the wave functions, thereby substantially reducing the 
computational cost and memory without loss of accuracy.

\subsection{Parallelization on Supercomputers}
We now describe our multi-level parallel implementation of the
multiterminal transport theory outlined above. First, the matrices are
distributed according to the two-dimensional block-cyclic data layout
scheme used by ScaLAPACK. Depending on the matrix size, one can use
$n\times n$ processors for matrix operations (typically n = 1 to 4 in
our applications). Second, parallelization proceeds over the energy
points used in the integration to obtain the charge density
matrix. Third, potentials and density matrices are also parallelized
over the 3D processor grid $pe_x, pe_y, pe_z$, where $pe_x\times
pe_y\times pe_z$ is the total number of processors. This step of
parallelization drastically accelerates the Poisson equation solver during
the self-consistent iterations.  Fourth, parallelization over the bias
points is trivial and can be achieved with nearly 100\% efficiency.

For the zero bias calculation, the computational cost for a multiterminal
system is about the same as that for a two-terminal system if the
number of atoms in the scattering region is the same. However, for the 
multiterminal system, an additional computational time is required for 
the nonequilibrium calculation. This is because, as the number
of leads increases, the number of terms in the density matrix also increases 
(see in Eq. 6). The most time-consuming part of the entire computation
is the matrix inversion needs to calculate the Green functions. This part 
scales nearly linearly if one takes into consideration the sparsity 
feature of the Hamiltonian and overlap matrices.

\section{Applications}
\label{sec:applications}
\subsection{Radialene molecule}
\label{subsec:radialene}
\begin{figure}
  \centering
  \includegraphics[width = 0.8\columnwidth]{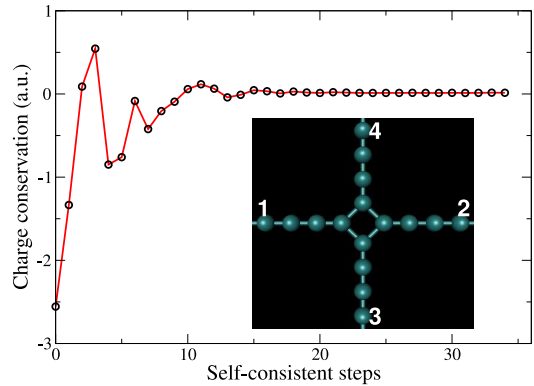}
  \caption{(Color online) Charge convergence of the radialene system
   with SC steps at zero bias. In the inset, a schematic 
   view of the central region of the four-terminal radialene junction is 
   shown, with the number 1, 2, 3 and 4 marking the positions of the leads.
   The size of the system is 60 Bohr $\times$ 60 Bohr.}
  \label{fig:sahafig4}
\end{figure}

In order to illustrate the proposed approach for calculating 
the conductance of a multiterminal molecular device,  
we choose a four-terminal junction of radialene molecule connected 
to semi-infinite carbon chains as a first example. A schematic diagram is shown in the inset of 
Fig.~\ref{fig:sahafig4}. The system has C$_{4v}$ symmetry. Applying our 
technique to this system, we expect to see the same symmetry in the 
converged potential profile, which should also be reflected in the 
transmission curves.

Our nonequilibrium Green function (NEGF) technique 
\cite{Larade01, Brandbyge02} uses a basis of optimal localized 
orbitals. \cite{Fattebert00, Nardelli01} The atom-centered orbitals
are optimized variationally in the equilibrium geometry. In the
radialene based system, we include 48 atoms in the calculation and 
each atom has six orbitals with the radii of 9 Bohr. 
A self-consistent calculation is carried out within an extended 
zone around the scattering region. For this system, the total charge
is converged after 18 steps of the SC process, as shown 
in Fig.~\ref{fig:sahafig4}. The charge density determines the potential.

\begin{figure}
  \centering
  \includegraphics[width = 0.9\columnwidth]{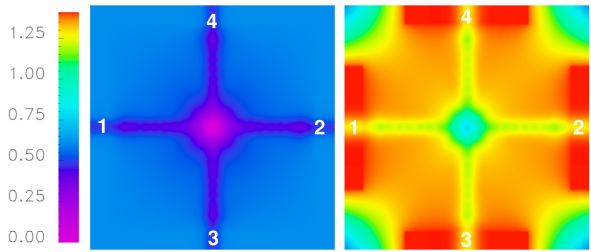}
  \caption{(Color online) (Left) Self-consistent converged potential 
   profile (same as system size, i.e., 60 Bohr $\times$ 60 Bohr) 
   at zero bias of the four-terminal radialene system, plotted 
   4.88 Bohr above the atomic plane. (Right) The converged potential 
   profile after applying an identical bias at $0.8 \,V$ through all the leads.
   Both the images are shown in same color scale (in eV), to compare 
   the relative heights of the potentials.} 
  \label{fig:sahafig5}
\end{figure}

Fig.~\ref{fig:sahafig5} (left) shows the converged potential profile at 
zero bias 4.88 Bohr above the atomic plane. The C$_{4v}$ symmetry of the
system is reflected in its potential profile. After the convergence of the charge
density is achieved for the equilibrium density matrix, we apply
the bias voltage through the leads. At this stage, the nonequilibrium 
part of the density matrix (see Eq.~\ref{eq:10}) is included through an
iterative process. In Fig.~\ref{fig:sahafig5} (right) we show the converged
potential profile after applying an identical bias of $0.8 \,V$ through
all the leads. It again shows the four-fold symmetry as expected.
It also shows that after the convergence, the potential matches very 
well at the boundary of the leads and the central molecule. 

\begin{figure}
  \centering
  \includegraphics[width = 0.9\columnwidth]{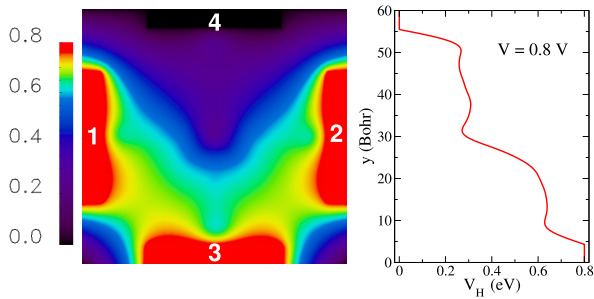}
  \caption{(Color online) (Left) Converged potential profile of the
   radialene system with a non-uniform bias voltage. (The color
   scale is in eV.) An identical bias of $0.8 \,V$ is applied 
   through three of the leads (denoted by 1, 2, 3) and no bias is 
   applied through the fourth lead. (Right) Potential drop along the 
   central line between the leads 3 and 4.} 
  \label{fig:sahafig6}
\end{figure}

We now examine the potential drop when different biases through the 
leads are applied. Here, a bias of $0.8 \,V$ is applied at the leads
$\region{L_1, L_2, L_3}$, while the fourth lead $\region{L_4}$ is 
at zero bias. After solving the Poisson equation, we observe a
uniform potential drop between the leads $\region{L_3}$ and
$\region{L_4}$ (see Fig.~\ref{fig:sahafig6} (left)). For testing 
purposes, we have also plotted the Hartree potential $V_H$ along 
the line connecting leads $\region{L_3}$ and $\region{L_4}$ in 
Fig.~\ref{fig:sahafig6} (right).

\begin{figure}
  \centering
  \includegraphics[width = 0.9\columnwidth]{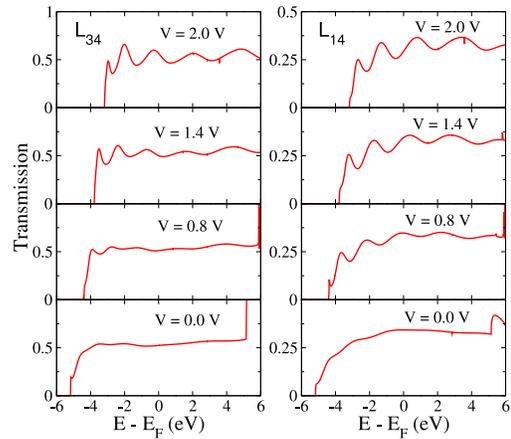}
  \caption{(Color online) Transmission curves of the four-terminal 
   radialene system with different bias voltages. The $E_F$ is
the Fermi energy of the lead 4. The bias geometry is the same
   as in Fig.~\ref{fig:sahafig6}. The left and right panels show transmission 
   through the leads $\region{L}_{34}$ and $\region{L}_{14}$, respectively.}
  \label{fig:sahafig7}
\end{figure}

Once the potential profile is self-consistently determined, the
transmission spectrum under the applied bias $V$ is calculated using
the Eq.~\ref{eq:transmission}. The transmission curves, shown in
Fig.~\ref{fig:sahafig7}, are computed for several bias voltages, with
the voltages being the same at $\region{L}_{1}, \region{L}_{2}$ and
$\region{L}_{3}$, while $\region{L}_{4}$ is being kept at $V = 0$.
The left and right panels in Fig.~\ref{fig:sahafig7} show the
transmission $\region{L}_{34}$ and $\region{L}_{14}$,
respectively. The carbon-atoms in the lead are fixed at equidistant
bond length. It follows that the lead has metallic character and
therefore no gap appears in the transmission curve (as would be the
case if the chain had been subjected to a Jahn-Teller
distortion). With zero bias, the transmission curve around the Fermi
level is almost constant. This nature of transmission is expected
because the free-electron-like $sp$-states of carbon contribute to the
transmission. However, with increasing bias, the transmission curve
starts to oscillate.  The origin of the oscillation is related to
  the choice of electrode. Here we used a simple, very idealized
  carbon chain made up of 8 atoms per lead. In order to examine the
  charge convergence in the scattering region, one needs to define a
  potential box around the molecule, where the Poisson equation is
  solved. However, because of the small number of carbon atoms in the
  lead, the potential box needs to be quite large, encompassing major
  parts of the leads. This creates finite-size effects, which are
  reflected in the transmissions showing oscillations.  As the bias
  is increased, the finite size effects also increase and this is why
  there are more oscillations in the transmission curves with higher bias.
  Therefore, the these oscillations are an artifact of the small
  number of atoms in our ``test'' leads. However, realistic
  molecular systems, with either thicker nanowires or bulk
  surfaces, do not show these artifacts.  Our investigations of
  two-terminal systems \cite{Lu05, Wang06, Ribeiro08} further
  demonstrate this claim. In addition, our ongoing investigations on
  more realistic four-terminal molecular junctions (to be communicated
  soon) are free of such oscillations.

Note that because of the C$_{4v}$ symmetry of the system, the transmissions 
$\region{L_{12}}$ and $\region{L_{34}}$ at zero bias are identical (not shown
in the figure). For the same reason, the transmission contributions
$\region{L_{13}}$, $\region{L_{14}}$, $\region{L_{23}}$ and
$\region{L_{24}}$ are equivalent (not shown in figure).

\begin{figure}
  \centering
  \includegraphics[width = 0.8\columnwidth]{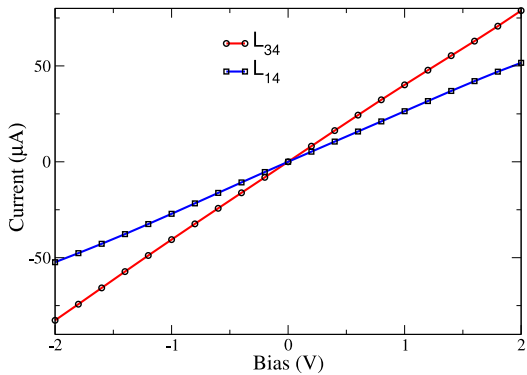}
  \caption{(Color online) Current-voltage characteristics of the
  four-terminal radialene junction. The bias geometry is as shown in 
  Fig.~\ref{fig:sahafig6}. The current contributions through
  the leads $\region{L}_{34}$, $\region{L}_{14}$ are displayed.} 
  \label{fig:sahafig8}
\end{figure}

We have also computed the I-V curves for this system, see
Fig.~\ref{fig:sahafig8}. The current contributions through leads
$\region{L_{34}}$ (red line) and $\region{L_{14}}$ (blue line) are
obtained in the voltage window $\pm2 \,V$. Both curves are increasing 
almost linearly because of the constant transmission around the
Fermi energy.

\subsection{Crossed carbon chains}

\begin{figure}
  \centering
   \includegraphics[width = 0.50\columnwidth]{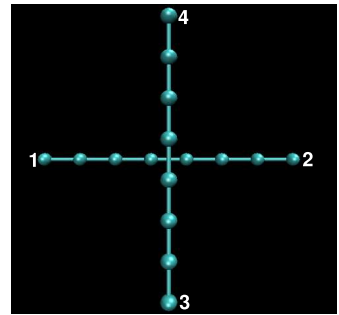}
   \caption{(Color online) A schematic diagram of the crossed-carbon-chains
    system. Three cases are considered, with distances 
    between the chains of 7.5, 5.0 and 2.5 Bohr.}  
  \label{fig:sahafig9}
\end{figure}

As a second example, we have chosen a four-terminal system consisting
of two crossed carbon chains. Our objective is to vertically bring the 
carbon chains closer and closer, and to see how the current varies with the
distance between the chains, possibly leading to a crossover from low to 
high coupling. In our study, we consider three distances
between the chains, $d =$ 7.5, 5.0 and 2.5 Bohr. To build the system,
we include a total of 66 carbon atoms, with 33 atoms in each chain.
Every atom has 6 basis orbitals with the radius of 9 Bohr, as in 
previous example. A schematic diagram of the system is shown in 
Fig.~\ref{fig:sahafig9}.

\begin{figure}
  \centering
  \includegraphics[width = 0.85\columnwidth]{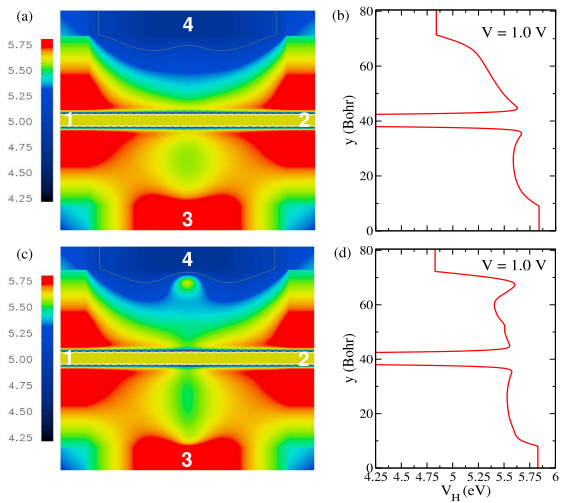}
  \caption{(Color online) (a) and (c) Comparison of the converged 
   potential profiles of the crossed-carbon-chains system when the distances 
   between the two carbon chains are 7.5 and 5.0 Bohr. In both the cases, we 
   apply same bias $0.5 \,V$ through the leads $\region{L_1, L_2}$, and ${L_3}$ 
   and bias $-0.5 \,V$ through the fourth lead $\region{L_4}$. The plotting
   passing through one of the chains. (b) and (d) Potential drops between 
   the leads $\region{L_3}$ to $\region{L_4}$ for (a) and (c) case.}
  \label{fig:sahafig10}
\end{figure}

After the charge convergence is achieved, we apply a $0.5 \,V$
bias through leads $\region{L_1, L_2, L_3}$ and a $-0.5 \,V$ bias
through the fourth lead $\region{L_4}$. The converged potential and
the charge density of the equilibrium system (i.\,e., at zero bias) is
used as the initial guess for the convergence of the new
nonequilibrium system. Fig.~\ref{fig:sahafig10} (left-top and bottom)
shows the converged potential profiles of the system when the
distances between the two carbon chains are 7.5 and 5.0 Bohr,
respectively. The plotting plane is parallel to the chains and passes
through one of them. Both figures are symmetric about the line connecting 
leads $\region{L_3}$ and $\region{L_4}$. To observe the potential drop 
along this line more clearly, we plot the Hartree potentials as shown 
in Fig.~\ref{fig:sahafig10} [(b) and (d)]. If the two chains
are sufficiently away from each other, e.\,g., separated by 7.5 Bohr,
the potential drop between two leads is smooth, and thus electron
tunneling along a given chain will be the largest in this case. As we
bring the chains closer, e.\,g., to a distance of 5.0 Bohr, the
probability for an electron to tunnel from one chain to the other
increases significantly. 

\begin{figure}
  \centering
  \includegraphics[width = 0.90\columnwidth]{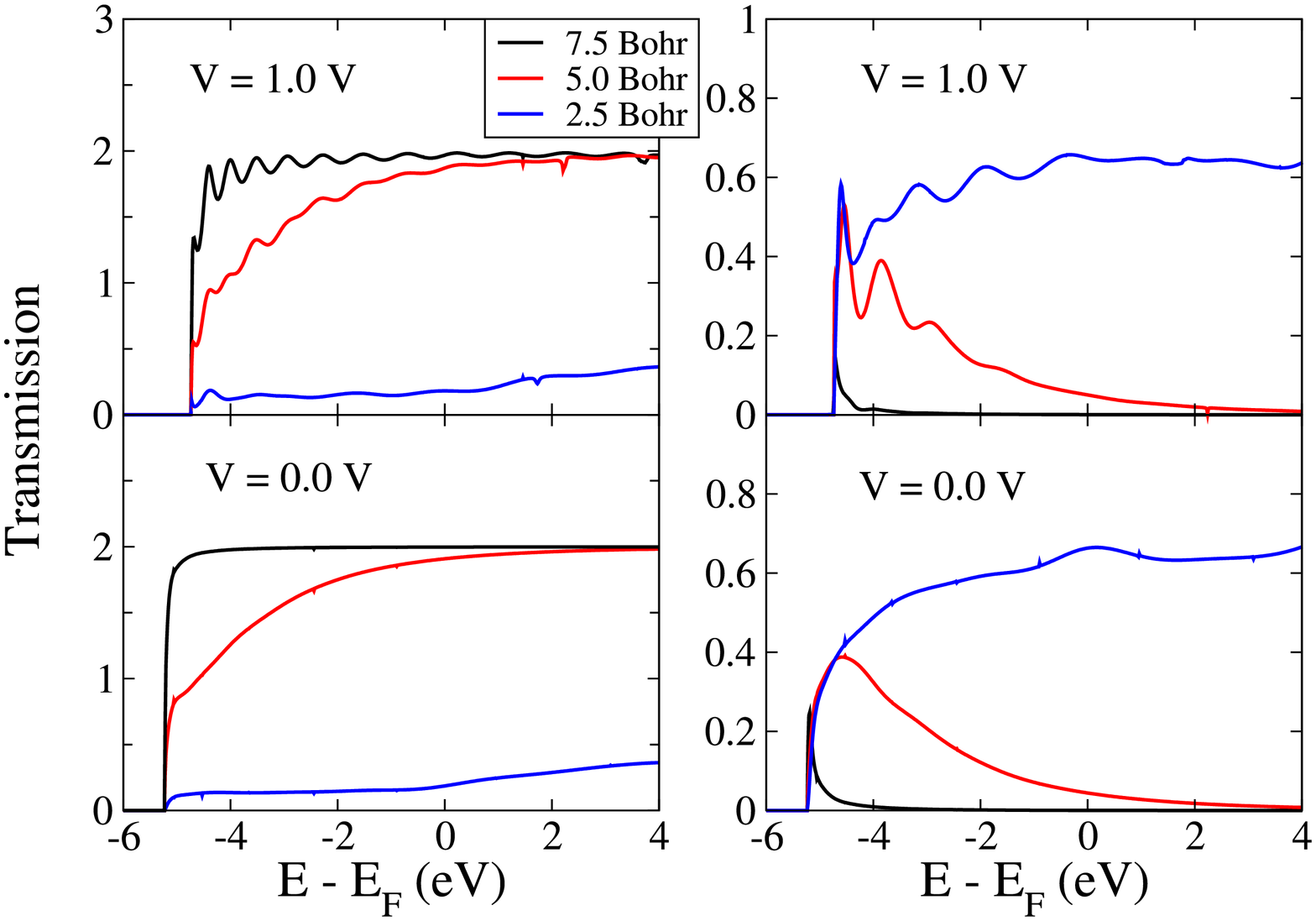}
  \caption{(Color online) Comparison of transmission curves in all three
   cases, where the distances between the chains are 7.5, 5.0 and 2.5 Bohr, 
   of the crossed-carbon-chain system with zero and non-zero biases (as in 
   Fig.~\ref{fig:sahafig10}). The left and right panels show transmission through 
   leads $\region{L_{34}}$ and $\region{L_{14}}$, respectively.}
  \label{fig:sahafig11}
\end{figure}

Fig.~\ref{fig:sahafig11} shows the transmission curves in all three cases
computed through leads $\region{L_{34}}$ (left panel) and
$\region{L_{14}}$ (right panel) with zero and non-zero biases. We
applied the same bias as in Fig.~\ref{fig:sahafig10}. The transmission of a
system consisting of a single ideal lead must be equal to the 
number of scattering channels at the energy $E$.\cite{Henk06, Saha08} 
If the carbon chains are far away from each 
another, e.\,g., at a distance of 7.5 Bohr, the overlap integral
between orbitals on atoms situated in two different leads is close to
zero. Therefore, each lead in the system will tunnel current as
an isolated electrode. This is why we observe a constant transmission
through $\region{L_{34}}$, while the transmission through 
$\region{L_{14}}$ is almost zero, as expected. In a carbon chain system, 
a two-fold degenerate band crosses the Fermi level. Consequently, there 
are two scattering channels and $G=2$.
As we decrease the distance between the chains,
to 5.0 and 2.5 Bohr, the orbital overlaps between
the chains become larger and larger.  This results in a decrease 
in transmission $\region{L_{34}}$ and increase
for $\region{L_{14}}$. The total transmission that
would occur only through $\region{L}_{34}$ in the bare lead case, is now
partially distributed to the other channels. At a finite bias, 
the transmission curves start to oscillate. As explained in the 
Subsection~\ref{subsec:radialene}, these oscillations are due  
to electrons that are attracted to the central region, 
thereby conserving the static charge and creating an ``electron-in-a-box''
effect with corresponding standing-wave-like oscillations.

\begin{figure}
  \centering
  \includegraphics[width =0.80\columnwidth]{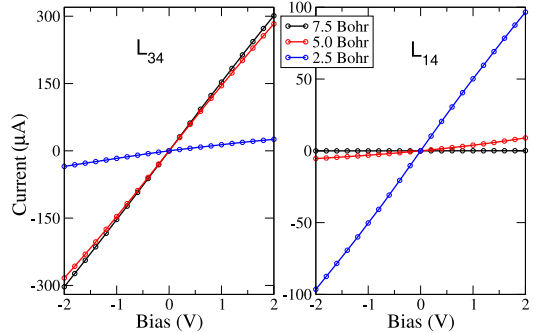}
  \caption{(Color online) Comparison of I-V curves with varying distances 
   between the chains of the crossed-carbon-chain system. The left and right 
   panels show the current contributions through leads $\region{L_{34}}$ 
   and $\region{L_{14}}$, respectively, with the bias voltage.} 
  \label{fig:sahafig12}
\end{figure}

Fig.~\ref{fig:sahafig12} shows the flow of current through the channels
$\region{L_{34}}$ (left panel) and $\region{L_{14}}$ (right panel).
We apply bias $V/2$ through the leads $\region{L_1, L_2, L_3}$ and
bias $-V/2$ through $\region{L_4}$, and compute the current for
distances 7.5, 5.0 and 2.5 Bohr between the carbon chains. In all
cases, as the bias increases, the current increases almost linearly. 
However, the nonlinearity is expected to be larger in a
semiconducting system. It should be noticed that the current
contribution through the $\region{L_{34}}$ channel increases with an
increase of the distance between the chains, while 
the current flow through the $\region{L_{14}}$ channel decreases. 
Note that the currents through the $\region{L_{14}}$ and
$\region{L_{24}}$ channels are identical because of the symmetry of the
system. It is interesting to see how the current flow
through a given channel, say $\region{L_{14}}$, decays
while moving the chains away from each other. We have
fitted the current contributions at a bias of $1.2 \,V$ as $I(d) = I_0 \
e^{-\beta d}$ and found the exponential decay constant $\beta$ to be 
0.97 Bohr$^{-1}$. 

\section{Summary and Conclusions}
\label{sec:concluding-remarks}
A new generalized approach for computing nonequilibrium quantum
transport in multiterminal systems from first principles is
developed within the framework of Keldysh theory. This advance opens
up new opportunities to study and design molecule-based electronic
devices. All calculations are performed at the density functional
theory level with full self-consistency under applied bias.  For
computational efficiency, we use a compact atom-centered optimized
orbitals obtained with a linear scaling method (N is the number of electrons)
for computing the electronic properties of the lead and the central
region. This basis is used to expand the Green functions, the
transmission function, and the charge density under bias, which are
self-consistently determined via contour integration. The
methodology is developed to scale well on massively-parallel
computers, and should therefore be applicable to systems of
realistic sizes.

  To demonstrate the suitability of the new technique for studying
  electron transport in multiterminal junctions, we have chosen two
  very simple four-terminal systems as test applications. In the first 
  example, a radialene system having $C_{4v}$ symmetry is used to test the 
  conservation of symmetry and the numerical robustness of our 
  implementation. In the second example, we have examined the conductance
  properties of two crossed carbon chains. The I-V characteristics of
  the chains show the expected trends with the changing strength of
  interactions. These demonstrations establish the general applicability 
  of the method. Since our code is efficient
  and highly parallel, we are able to deal with rather large systems.
  One such application (a four-terminal system consisting of an organic 
  molecule [9,10-Bis(($2'$-para-mercaptophenyl)-ethinyl)-anthracene] 
  connected to four gold nanowires) will be published elsewhere\cite{saha2009}.

\section*{ACKNOWLEDGMENTS}

Portions of this research was sponsored by the Laboratory Directed
Research and Development Program of Oak Ridge National Laboratory
(ORNL), managed by UT-Battelle, LLC for the U. S. Department of Energy
under Contract No. De-AC05-00OR22725 (KKS and VM), by DOE grants
DE-FG02-03ER46095 and DE-FG02-98ER45685, and by ONR grant
N000140610173 (WL and JB).
%
%

\end{document}